\documentclass[twocolumn,amsmath,amssymb]{revtex4}
\usepackage{graphicx}
\usepackage{dcolumn}
\usepackage{bm}

\begin{document}
\title[]
{Wormholes with a space- and time-dependent equation of state}
\author{Peter K.\,F. Kuhfittig}
\address{Department of Mathematics\\
Milwaukee School of Engineering\\
Milwaukee, Wisconsin 53202-3109}
\date{\today}

\begin{abstract}\noindent
The discovery that the Universe is undergoing an accelerated
expansion has suggested the existence of an evolving equation
of state.  This paper discusses various wormhole solutions in
a spherically symmetric spacetime with an equation of state
that is both space and time dependent.  The solutions obtained
are exact and generalize earlier results on static wormholes
supported by phantom energy.
\end{abstract}

\maketitle

PAC numbers: 04.20.Jb, 04.20.Gz


\section{Introduction}\noindent
Traversable wormholes, whose possible existence was first conjectured
by Morris and Thorne in 1988 \cite{MT88}, are actually shortcuts that
could in principle be used for traveling to remote parts of our
Universe or to different universes altogether.  The meticulous analysis
in \cite{MT88} has shown that such wormholes can only be held open by
the use of \emph{exotic matter}.  Such matter violates the weak energy
condition (WEC), which requires the stress-energy tensor
$T_{\alpha\beta}$ to obey
$T_{\alpha\beta}\mu^{\alpha}\mu^{\beta}\ge 0$ for all time-like
vectors and, by continuity, all null vectors.  For example, given the
radial outgoing null vector $(1,1,0,0)$, we obtain
\[
    T_{\hat{\alpha}\hat{\beta}}\mu^{\hat{\alpha}}\mu^{\hat{\beta}}
      =\rho+p\ge 0.
\]
(Recall that $T_{\hat{t}\hat{t}}=\rho$ and $T_{\hat{r}\hat{r}}=p$ in
the orthonormal frame of reference.)  So if the WEC is violated, we
have $\rho+p<0.$

Interest in traversable wormholes has increased in recent years due
to an unexpected connection, the discovery that our Universe is
undergoing an accelerated expansion \cite{aR98, sP99}.  This
acceleration, caused by a negative pressure \emph{dark energy},
implies that $\overset{..}{a}>0$ in the Friedmann equation
$\frac{\overset{..}{a}}{a}=-\frac{4\pi}{3}(\rho+3p).$  (Our units are
taken to be those in which $G=c=1.$)  The equation of state is $p=
-w\rho$, $w>\frac{1}{3}$, and $\rho>0.$  While the condition
$w>\frac{1}{3}$ is required for an accelerated expansion, larger
values for $w$ are also of interest.  For example, $w=1$ corresponds
to the existence of Einstein's cosmological constant.  (The usual
form of the equation of state is $p=w\rho$, $w<-\frac{1}{3}$, but
in this paper a positive parameter is computationally more
convenient.)

Of particular importance for us is the case $w>1$, referred to as
\emph{phantom energy}.  For this case we have $\rho+p<0,$ in violation
of the weak energy condition.  As noted realier, this condition is the
primary prerequisite for the existence of traversable wormholes.
(Strictly speaking, the notion of dark or phantom energy applies only
to a homogeneous distribution of matter in the Universe, while
wormhole spacetimes are necessarily inhomogeneous.  However, the
extension to spherically symmetric inhomogeneous spacetimes has been
carried out \cite{sS05}.)

Two recent papers \cite{fL07, fR07} discussed wormhole solutions
that depend on a variable equation of state, i. e.,
$\frac{p}{\rho}=-m(r),$ where $m(r)>1$ for all $r$, corresponding
to a variable $w$.  The variable $r$ refers to the radial
coordinate in the line element
\begin{equation}\label{E:line1}
  ds^2=-e^{2f(r)}dt^2+\frac{1}{1-b(r)/r}dr^2
       +r^2(d\theta^2+\text{sin}^2\theta\,d\phi^2);
\end{equation}
in other words, $m=m(r)$ is independent of direction.  It is shown
in Ref. \cite{fR07} that given a specific \emph{shape function}
$b=b(r)$, it is possible to determine $m=m(r)$ and vice versa.
It is also assumed that for the \emph{redshift function} $f(r)$,
$f'(r)\equiv 0$, referred to as the ``zero tidal-force solution"
in Ref.~\cite{MT88}.

An earlier study \cite{BS94} assumed that the equation of state is a
function of time.  In this paper we will assume that $w=w(r,t)$ is a
continuous function of both $r$ and $t$, so that the equation of
state is $p=-w(r,t)\rho$, $w(r,t)>1$.  As in Ref.~\cite{fR07},
however, we retain the assumption that the function values are
independent of direction.  For reasons that will become apparent
later, we also assume that the change in $t$ is very gradual.

According to Ref. \cite{aV05}, recent data from supernovae, CMB,
and weak gravitational lensing favor an evolving equation of state,
possibly even with $w>1$.  These findings provided the motivation
for discussing phantom stars in Ref. \cite{DGL08} and
traversable wormholes in this work.  Evolving wormhole geometries
are also discussed in Refs. \cite{lA98} and \cite{KS96}.

The main goal in this paper is to show that the time-dependent metric
describes a slowly evolving wormhole structure without assigning
specific functions to $b$ and $w$.  Moreover, the function $f$ in
line element (\ref{E:line1}) need not be a constant.  All the
solutions obtained are exact and generalize earlier results on
static wormholes supported by phantom energy.

\section{The shape function and flare-out condition}\noindent
Normally, one would begin with the general line element
\begin{equation}\label{E:line2}
  ds^2=-e^{2\gamma(r)}dt^2+e^{2\alpha(r)}dr^2+r^2(d\theta^2
     +\text{sin}^2\theta\,d\phi^2).
\end{equation}
In view of line element (\ref{E:line1}),
\[
   e^{2\alpha(r)}=\frac{1}{1-b(r)/r}.
\]
As already noted, $b=b(r)$ is the \emph{shape function};
$b(r_0)=r_0$, where $r=r_0$ is the radius of the throat.  The
shape function must satisfy the flare-out condition $b'(r_0)<1$,
a consequence of the WEC violation.  Observe also that
\[
   \lim_{r \to r_0+}\alpha(r)=+\infty.
\]
Recall that $\gamma(r)$ is referred to as the \emph{redshift
function}.  This function must be finite everywhere to avoid an event
horizon.

Returning to the function $w=w(r,t)$, since, for any fixed $t$,
the function is invariant under rotation, we need a time-dependent
metric with  the same property \cite{pK02}:
\begin{equation}\label{E:line3}
   ds^2 =-e^{2\gamma(r,t)}dt^2+e^{2\alpha(r,t)}dr^2+r^2(d\theta^2+
      \text{sin}^2\theta\, d\phi^2).
\end{equation}
This metric describes a spherically symmetric evolving wormhole.
Observe that the shape function is now given by
\begin{equation}\label{E:shape1}
    b(r,t)=r(1-e^{-2\alpha(r,t)}).
\end{equation}

To obtain a traversable wormhole, the shape function must not
only obey the usual flare-out condition at the throat, but must
be carefully modified to accommodate the time dependence.  To see
how, let us introduce the time-dependent sphere $r=r_t$, the
analogue of $r=r_0$, subject to the following condition:
\begin{equation}\label{E:flare}
  b(r_t,t)=r_t\quad \text{and} \quad \frac{\partial}{\partial r}
       b(r_t,t)<1 \quad \text{for all}\,\,t.
\end{equation}
Unfortunately, however, the sphere $r=r_t$ is the \emph{center}
of the wormhole, not the throat.  In fact, according to Hochberg
and Visser \cite{HV98}, in a time-dependent wormhole spacetime,
there are actually two throats of instantaneous radii $r=r_1$
and $r=r_2$, respectively, on opposite sides of the center.
A throat, such as $r=r_1$, is located entirely within one
time-slice and constitutes a hypersurface of minimal area.
Thanks to the spherical symmetry, the throat is therefore
another sphere with $r_1>r_t$.  At this point we need to
make an additional assumption: for any fixed $t$,
$b=b(r,t)$ must be a typical shape function, increasing and
concave down, at least near the center, so that the slope
continues to decrease in the outward radial direction.
(See, for example, the discussion of profile curves and
embedding diagrams in Ref. \cite{dD01}). Then  the condition
$(\partial/\partial r)b(r_t,t)<1$ will automatically result
in $(\partial/\partial r)b(r_1,t)<1$ and
$(\partial/\partial r)b(r_2,t)<1$.  Having now made sure that
the flare-out conditions are satisfied, we can safely write
our exact solutions in terms of the center $r=r_t$.  (The
connection to the WEC violation will be discussed in the
next section.)

A final requirement, discussed later, is asymptotic flatness:
$b(r,t)/r\rightarrow 0$ as $r\rightarrow\infty.$

\section{The Einstein tensor and equation of state}
\noindent
The components of the Einstein tensor in the orthonormal frame are
available from Ref. \cite{pK02}:
\begin{equation}\label{E:Einstein1}
   G_{\hat{t}\hat{t}}=\frac{2}{r}e^{-2\alpha(r,t)}\frac{\partial}
   {\partial r}\alpha(r,t)+\frac{1}{r^2}(1-e^{-2\alpha(r,t)}),
\end{equation}
\begin{equation}\label{E:Einstein2}
    G_{\hat{r}\hat{r}}=\frac{2}{r}e^{-2\alpha(r,t)}\frac{\partial}
   {\partial r}\gamma(r,t)-\frac{1}{r^2}(1-e^{-2\alpha(r,t)}),
\end{equation}
\begin{equation}\label{E:Einstein3}
   G_{\hat{t}\hat{r}}=\frac{2}{r}e^{-\gamma(r,t)}e^{-\alpha(r,t)}
   \frac{\partial}{\partial t}\alpha(r,t),
\end{equation}
\begin{multline}\label{E:Einstein4}
   G_{\hat{\theta}\hat{\theta}}=G_{\hat{\phi}\hat{\phi}}=
   -e^{-2\gamma(r,t)}\left[\frac{\partial^2}{\partial t^2}
    \alpha(r,t)
     \phantom{\left(\frac{\partial}{\partial t}\alpha(r,t)
    \right)^2}\right.\\
    \left.-\frac{\partial}{\partial t}\gamma(r,t)
    \frac{\partial}{\partial t}\alpha(r,t)
      \right.
     \left.+\left(\frac{\partial}{\partial t}\alpha(r,t)
    \right)^2\right]\\
     -e^{-2\alpha(r,t)}\left[-\frac{\partial^2}{\partial r^2}
    \gamma(r,t)
     \phantom{\left(\frac{\partial}{\partial t}\alpha(r,t)
    \right)^2}\right.\\
    \left.+\frac{\partial}{\partial r}\gamma(r,t)
    \frac{\partial}{\partial r}\alpha(r,t)
      \right.
     \left.-\left(\frac{\partial}{\partial r}\gamma(r,t)
    \right)^2\right]\\
  -\frac{1}{r}e^{-2\alpha(r,t)}\left(-\frac{\partial}{\partial r}
   \gamma(r,t)+\frac{\partial}{\partial r}\alpha(r,t)\right).
\end{multline}

Recall that from the Einstein field equations in the orthonormal frame,
$G_{\hat{\alpha}\hat{\beta}}=8\pi T_{\hat{\alpha}\hat{\beta}}$,
the components of the Einstein tensor are proportional to the
components of the stress-energy tensor.  In particular,
$T_{\hat{t}\hat{r}}=T_{\hat{r}\hat{t}}=\frac{1}{8\pi}
G_{\hat{t}\hat{r}}=\pm f$ is interpreted as the energy flux
in the outward radial direction \cite{tR93}.  The WEC now
becomes $\rho+p\pm 2f\ge 0.$ So if the WEC is violated, then
\begin{multline}\label{E:WEC}
  \frac{1}{8\pi}\left[\frac{2}{r}e^{-2\alpha(r,t)}\left(
   \frac{\partial}{\partial r}\alpha(r,t)+\frac{\partial}
   {\partial r}\gamma(r,t)\right)\right.\\
    \left.\pm\frac{4}{r}e^{-\gamma(r,t)}e^{-\alpha(r,t)}
     \frac{\partial}{\partial t}\alpha(r,t)\right]<0.
\end{multline}
It is easy to check that since $\lim_{r \to r_t+}\alpha(r,t)
=+\infty$ and $(\partial/\partial r)\gamma(r,t)$ is finite,
Eq. (\ref{E:flare}) implies that
$(\partial/\partial r)\alpha(r,t)+(\partial/\partial r)
\gamma(r,t)$ becomes large and negative as $r\rightarrow r_t+$.
Moreover, it will be seen in the next section that
$\alpha(r,t)$ depends directly on $w(r,t)$.  So if $w(r,t)$
changes slowly enough with respect to time, then the last
term on the left-hand side of inequality (\ref{E:WEC})
becomes negligible, ensuring that the WEC will always be
violated at $r=r_t$ and hence at the two throats.

The reliance on Eq. (\ref{E:flare}) to show the WEC violation
does not explain why the condition
$(\partial/\partial r)b(t_t,t)<1$ should hold in the first
place: this condition follows from the phantom-like equation
of state $p=-w(r,t)\rho$, $w(r,t)>1$, as we will see in the
next section.  So the WEC violation could also be expressed
as
\[
  \rho+p\pm 2f=\rho[1-w(r_t,t)]\pm 2f<0,
\]
leading to the same conclusion.

From the Einstein field equations
$G_{\hat{\alpha}\hat{\beta}}=8\pi T_{\hat{\alpha}\hat{\beta}}$
and the equation of state $p=-w(r,t)\rho,$ we have
$G_{\hat{t}\hat{t}}=8\pi\rho$ and $G_{\hat{r}\hat{r}}=
8\pi[-w(r,t)]\rho$.  Using Eqs. (\ref{E:Einstein1}) and
(\ref{E:Einstein2}), we obtain the following system of equations:
\begin{multline*}
   G_{\hat{t}\hat{t}}=8\pi T_{\hat{t}\hat{t}}=\frac{2}{r}
   e^{-2\alpha(r,t)}\frac{\partial}{\partial r}\alpha(r,t)\\
   +\frac{1}{r^2}\left(1-e^{-2\alpha(r,t)}\right),
\end{multline*}
\begin{multline*}
   G_{\hat{r}\hat{r}}=8\pi T_{\hat{r}\hat{r}}=8\pi[-w(r,t)]\rho\\
   =\frac{2}{r}e^{-2\alpha(r,t)}\frac{\partial}{\partial r}
   \gamma(r,t)-\frac{1}{r^2}\left(1-e^{-2\alpha(r,t)}\right).
\end{multline*}
After substituting and rearranging terms, we have
\begin{multline}\label{E:generalequation}
  w(r,t)\frac{\partial}{\partial r}\alpha(r,t)\\
   =-\frac{\partial}{\partial r}\gamma(r,t)
  -\frac{1}{2r}\left(e^{2\alpha(r,t)}-1 \right)[w(r,t)-1].
\end{multline}

\section{The redshift function-Exact solutions}
    \label{S:exact}\noindent
In this section we solve Eq.~(\ref{E:generalequation}) by letting
the redshift function take on a specific form. One obvious choice
is $\gamma(r,t)\equiv\,\text{constant}$, so that $\frac{\partial}
{\partial r}\gamma(r,t)\equiv 0;$ the other is $\frac{\partial}
{\partial r}\gamma(r,t)=\frac{w(r,t)-1}{2r}$, allowing the solution
of Eq.~(\ref{E:generalequation}) by separation of variables.

If $\frac{\partial}{\partial r}\gamma(r,t)\equiv 0,$ then Eq.~
(\ref{E:generalequation}) becomes
\[
  \frac{\partial}{\partial r}\alpha(r,t)=-\frac{1}{2r}
  \left(e^{2\alpha(r,t)}-1\right)\left(1-\frac{1}{w(r,t)}\right).
\]
Rewritten as
\[
\frac{2\frac{\partial}{\partial r}\alpha(r,t)}{e^{2\alpha(r,t)}-1}
  =-\frac{1}{r}\left(1-\frac{1}{w(r,t)}\right),
\]
one recognizes the form $\int\frac{du}{e^u-1}=\text{ln}
\,(e^u-1)-u.$  The result is
\[
   \text{ln}\left(e^{2\alpha(r,t)}-1\right)-2\alpha(r,t)=
  -\text{ln}\,r+\int^r_c\frac{dr'}{r'w(r',t)}.
\]
Next, we solve for $e^{2\alpha(r,t)}$:
\begin{equation}\label{E:alpha}
   e^{2\alpha(r,t)}=\left[1-\frac{e^{\int^r_cdr'/
   r'w(r',t)}}{r}\right]^{-1}.
\end{equation}
Evidently,
\begin{equation}
   b(r,t)=e^{\int^r_cdr'/r'w(r',t)}.
\end{equation}
At $r=r_t,$ we have
\begin{equation}\label{E:shapeatthroat}
   \frac{e^{\int^{r_t}_cdr/rw(r,t)}}{r_t}=1
\end{equation}
due to the requirement $b(r_t,t)=r_t$.  Also,
\begin{multline*}
  \left.\frac{\partial}{\partial r}b(r_t,t)=e^{\int^r_cdr'/
   r'w(r',t)}\frac{1}{rw(r,t)}\right|_{r=r_t}\\
   =r_t\frac{1}{r_tw(r_t,t)}=\frac{1}{w(r_t,t)}<1.
\end{multline*}
So the flare-out condition is met without any additional
assumptions on either $b$ or $w$.  Since $w=w(r,t)$ is
continuous, the constant $c$ is uniquely determined by
Eq.~(\ref{E:shapeatthroat}):
\[
  \int^{r_t}_c\frac{dr}{rw(r,t)}=\text{ln}\,r_t.
\]
So $c$ is actually a function of $t$.  As an
illustration, in the special case $w(r,t)\equiv K$, a constant,
we obtain $c=r_0^{1-K}$, since $r_t=r_0$ in the static case.
The result is
\[
   e^{2\alpha(r)}=\frac{1}{1-\left(r_0/r\right)^{1-1/K}},
\]
which is Lobo's solution \cite{fL05}.

The line element is now seen to be
\begin{multline}\label{E:line4}
  ds^2=-e^{2\gamma(r,t)}dt^2
   +\left[1-\frac{e^{\int^r_cdr'/r'w(r',t)}}
   {r}\right]^{-1}dr^2\\
    +r^2(d\theta^2+\text{sin}^2\theta\,d\phi^2).
\end{multline}

\emph{Remark:} Returning to inequality (\ref{E:WEC}), since
$\alpha(r,t)$ depends on $w(r,t)$ [see Eq. (\ref{E:alpha})],
$(\partial/\partial t) \alpha(r,t)$ is small only if
$(\partial/\partial t)w(r,t)$ is small, explaining our
earlier requirement that $w(r,t)$ change only gradually
 with respect to time.  In other words, a wormhole can
 only be sustained if the equation of state evolves
 sufficiently slowly.

For the other choice of $\gamma(r,t)$,
\[
   \frac{\partial}{\partial r}\gamma(r,t)=\frac{w(r,t)-1}{2r},
\]
Eq.~(\ref{E:generalequation}) becomes
\begin{multline*}
  w(r,t)\frac{\partial}{\partial r}\alpha(r,t)\\
  =-\frac{w(r,t)-1}{2r}-\frac{1}{2r}\left(e^{2\alpha(r,t)}-1
     \right)\left[w(r,t)-1\right]\\
   =\frac{1}{2r}\left[w(r,t)-1\right]
            \left(-1-e^{2\alpha(r,t)}+1\right),
\end{multline*}
or
\[
   \frac{2\frac{\partial}{\partial r}\alpha(r,t)}
   {e^{2\alpha(r,t)}}=-\frac{1}{r}\left(1-\frac{1}{w(r,t)}\right).
\]
Solving, we get
\[
   e^{-2\alpha(r,t)}=\int^r_c\frac{1}{r'}
      \left(1-\frac{1}{w(r',t)}\right)dr'.
\]
So by Eq.~(\ref{E:shape1}),
\begin{multline*}
   b(r,t)=r\left(1-e^{-2\alpha(r,t)}\right)\\
   =r\left[1-\int^r_c\frac{1}{r'}\left(1-\frac{1}{w(r',t)}
     \right)dr'\right].
\end{multline*}
It now becomes apparent that $c=r_t$ for all $t$, since we must
have $b(r_t,t)=r_t$.  Thus
\begin{equation}\label{E:shape2}
  b(r,t)=r\left[1-\int^r_{r_t}\frac{1}{r'}\left(1-\frac{1}
   {w(r',t)}\right)dr'\right],
\end{equation}
so that the line element becomes
\begin{multline}\label{E:line5}
  ds^2=-e^{2\gamma(r,t)}dt^2+
  \left[\int^r_{r_t}\frac{1}{r'}
  \left(1-\frac{1}{w(r',t)}\right)dr'\right]^{-1}dr^2\\
      +r^2(d\theta^2+\text{sin}^2\theta\,d\phi^2).
\end{multline}
Once again,
\[
  \frac{\partial}{\partial r}b(r_t,t)=\frac{1}{w(r_t,t)}<1.
\]
Returning to the redshift function, we have up to this point
\begin{equation}\label{E:redshift1}
  \gamma(r,t)=\int^r_{c_1}\frac{w(r',t)-1}{2r'}dr'.
\end{equation}
The constant $c_1$ will be obtained from the junction conditions,
described below.

Based on previous studies involving static wormholes supported by
phantom energy \cite{fL05, oZ05, pK06}, our spacetime is not likely
to be asymptotically flat.  The wormhole material will therefore
have to be cut off at some $r=a$ and joined to an external
Schwarzschild spacetime.  Moreover, in Ref.~\cite{pK06}, $b=b(r)$
actually attains a maximum value at some $r=a$, which then becomes
a natural place at which to perform the junction.  Accordingly,
we will assume that for any fixed $t$, $b(r,t)$ has a maximum value
at some $r=a$.  (While other values could be chosen, the choice
suggested here yields a particularly elegant solution.)  So we
proceed by determining the critical value for $t$ fixed:
\begin{multline*}
  \frac{\partial}{\partial r}b(r,t)=1-\int^r_{r_t}
  \frac{1}{r'}\left(1-\frac{1}{w(r',t)}\right)dr'\\
    +r\left[-\frac{1}{r}\left(1-\frac{1}{w(r,t)}\right)\right]=0.
\end{multline*}
By assumption, equality holds for some $r=a$.  As a consequence,
\begin{equation}\label{E:redshift2}
   \int^a_{r_t}\frac{1}{r}\left(1-\frac{1}{w(r,t)}\right)dr
       =\frac{1}{w(a,t)}.
\end{equation}

As noted in Ref. \cite{fL05}, to match our interior solution to
the exterior Schwarzschild solution
\begin{multline*}
  ds^2=-\left(1-\frac{2M}{r}\right)dt^2+
       \left(1-\frac{2M}{r}\right)^{-1}dr^2\\
           +r^2(d\theta^2+\text{sin}^2\theta\,d\phi^2)
\end{multline*}
at $r=a$ ($t$ fixed) requires continuity of the metric.  Because
of the assumption of spherical symmetry, the components
$g_{\hat{\theta}\hat{\theta}}$ and $g_{\hat{\phi}\hat{\phi}}$ are
already continuous \cite{fL05}.  As a result, the continuity
requirement has to be imposed only on the remaining components.
For every fixed $t$,
\[
  g_{\hat{t}\hat{t}(\text{int})}(a)=g_{\hat{t}\hat{t}(\text{ext})}(a)
    \quad\text{and}\quad
  g_{\hat{r}\hat{r}(\text{int})}(a)=g_{\hat{r}\hat{r}(\text{ext})}(a)
\]
for the interior and exterior components, respectively.  These
conditions now imply that (for every fixed $t$)
\[
  \gamma_{\text{int}}(a)=\gamma_{\text{ext}}(a)
    \quad\text{and}\quad
  b_{\text{int}}(a)=b_{\text{ext}}(a).
\]
Hence
\[
  e^{2\alpha(a,t)}=\frac{1}{1-\frac{b(a,t)}{a}}=\frac{1}{1-\frac{2M}{a}}.
\]
We now see that the total mass of the wormhole for $r\le a$ is
given by $M=\frac{1}{2}b(a,t)$ for every fixed $t$.  So by Eqs.
(\ref{E:shape2}) and (\ref{E:redshift2})
\begin{equation}\label{E:mass}
   M=\frac{1}{2}a\left(1-\frac{1}{w(a,t)}\right).
\end{equation}
By Eq.~(\ref{E:redshift1}),
\begin{multline*}
  e^{2\gamma(a,t)}=e^{\int^a_{c_1}(w(r,t)-1)/rdr}
       =1-\frac{2M}{a}\\
  =1-\frac{2}{a}\cdot\frac{1}{2}a\left(1-\frac{1}{w(a,t)}\right)
\end{multline*}
or
\begin{equation}\label{E:elegant}
  e^{2\gamma(a,t)}=\frac{1}{w(a,t)}.
\end{equation}
Eq.~(\ref{E:elegant}) can now be used to determine $c_1=c_1(t),$
thereby completing the line element, Eq.~(\ref{E:line5}).

As a concrete illustration of this procedure, if
$w(r,t)\equiv K$, then $r_t=a_0$ again.  Then from Eqs.
(\ref{E:mass}) and (\ref{E:shape2}) we deduce that
$a=r_0e^{1/(K-1)}$.  Eqs. (\ref{E:elegant}) and
(\ref{E:redshift1}) then yield $c_1=r_0(Ke)^{1/(K-1)}$.

\section{Additional solutions}
As noted in Ref.~\cite{pK06}, to obtain additional exact
solutions, $\gamma$ must depend directly on $\alpha$.  The
corresponding condition for the time-dependent case is
\begin{equation}\label{E:redshift3}
   \frac{\partial}{\partial r}\gamma(r,t)=F[\alpha(r,t)]
      \frac{\partial}{\partial r}\alpha(r,t)
\end{equation}
for some elementary function $F$.  Since these cases are just
extensions of the cases discussed in Ref.~\cite{pK06}, we will
merely summarize the results.

If
\begin{equation}\label{E:redshift4}
  F[\alpha(r,t)]=-\frac{w(r,t)}{e^{2\alpha(r,t)}},
\end{equation}
then Eq.~(\ref{E:generalequation}) becomes
\begin{multline*}
  w(r,t)\frac{\partial}{\partial r}\alpha(r,t)\\
   =\frac{w(r,t)\frac{\partial}{\partial r}\alpha(r,t)}
      {e^{2\alpha(r,t)}}
          -\frac{1}{2r}\left(e^{2\alpha(r,t)}-1\right)
              [w(r,t)-1].
\end{multline*}
The solution is
\[
   e^{2\alpha(r,t)}=\left[\int^r_{r_t}\frac{1}{r'}
     \left(1-\frac{1}{w(r',t)}\right)dr'\right]^{-1}.
\]
(If $w(r,t)\equiv K$, the solution reduces to
\[
  e^{2\alpha(r)}=\frac{1}{\text{ln}\left(r/r_0\right)
    ^{1-1/K}},
\]
discussed in Ref.~\cite{pK06}.)  Also,
\[
  b(r,t)=r\left[1-\int^r_{r_t}\frac{1}{r'}
     \left(1-\frac{1}{w(r',t)}\right)dr'\right].
\]
As in the previous cases,
\[
  b(r_t,t)=r_t \quad \text{and} \quad \frac{\partial}{\partial r}
     b(r_t,t)=\frac{1}{w(r_t,t)}<1.
\]

The determination of the redshift function and the junction to an
exterior Schwarzschild solution follow along the lines discussed
in Sec.~\ref{S:exact}.

Another solution comes from
\begin{equation}\label{E:redshift5}
  F[\alpha(r,t)]=-\frac{2w(r,t)}{e^{2\alpha(r,t)}+1}.
\end{equation}\underline{}
Substituting in Eq.~(\ref{E:generalequation}) and simplifying, we
get
\begin{equation*}
  \frac{\frac{\partial}{\partial r}\alpha(r,t)}
    {e^{2\alpha(r,t)}-1}+\frac{1}{2r}
        \left(1-\frac{1}{w(r,t)}\right)
    =\frac{2\frac{\partial}{\partial r}\alpha(r,t)}
         {e^{4\alpha(r,t)}-1}.
\end{equation*}
The solution is
\[
   e^{2\alpha(r,t)}=\left[e^{\int^r_{r_t}(1/r')
      \left(1-1/w(r',t)\right)dr'}-1\right]^{-1}.
\]
(If $w(r,t)\equiv K$, this reduces to
\[
  e^{2\alpha(r)}=\frac{1}{\left(r/r_0\right)^{1-1/K}-1},
\]
also discussed in Ref.~\cite{pK06}.)  Here
\begin{multline*}
  b(r,t)=r\left(1-e^{-2\alpha(r,t)}\right)\\
   =r\left(2-e^{\int^r_{r_t}(1/r')
         \left(1-1/w(r',t)\right)dr'}\right).
\end{multline*}
Once again,
\[
  b(r_t,t)=r_t \quad \text{and} \quad \frac{\partial}{\partial r}
    b(r_t,t)=\frac{1}{w(r_t,t)}<1.
\]

\section{Discussion}\noindent
Recent astrophysical observations from supernovae, CMB, and weak
gravitational lensing have suggested that the equation of state
not only evolves but actually favors a value of $w$ in the
phantom-energy range.  These observations have provided a strong
motivation for studying traversable wormholes with an evolving
equation of state.  This paper discusses several exact solutions
of the Einstein field equations describing traversable wormholes
supported by a generalized form of phantom energy: the evolving
equation of state is given by $p=-w(r,t)\rho$, $w(r,t)>1$.  The
function $w=w(r,t)$ is a continuous function of $r$ and $t$,
invariant under rotation.  Such wormholes can only be sustained
if the equation of state evolves sufficiently slowly.

Since we are dealing with a variable equation of state, $w(r,t)$ 
could have changed from $w<1$ to $w>1$, sometimes referred to as 
``crossing the phantom divide."  According to the detailed analysis 
in Ref. \cite{DGL08}, this could have resulted in a topology change.  
In other words, it is conceivable that a star could become a wormhole.

In a different scenario \cite{pK08}, if $w(r,t)$ had crossed the 
phantom divide some time in the past, wormholes could have formed 
spontaneously.  Assuming that the present dark-energy phase is best 
modeled after Einstein's cosmological constant \cite{rB08}, these 
wormholes would have formed event horizons (to become black holes) 
or quasihorizons.  In the latter case, they would still be wormholes 
but with enormous surface stresses.

\end{document}